\documentclass[articel]{aa}
\usepackage[varg]{txfonts}
\begin{document}
\title{The lost sunspot cycle: New support from $^{10}$Be measurements}
\author{C.~Karoff\inst{1,2}
  \and F.~Inceoglu\inst{2} 
  \and M.~F.~Knudsen\inst{1}
  \and J.~Olsen\inst{3}
  \and A.~Fogtmann-Schulz\inst{1}}
\offprints{C. Karoff, \email{karoff@phys.au.dk}}
\institute{Department of Geoscience, Aarhus University, H{\o}egh-Guldbergs Gade 2, DK-8000 Aarhus C, Denmark
  \and Stellar Astrophysics Centre, Department of Physics and Astronomy, Aarhus University, Ny Munkegade 120, DK-8000 Aarhus C,
Denmark
  \and AMS, 14C Dating Centre, Department of Physics, Aarhus University, Ny Munkegade 120, DK-8000 Aarhus C, Denmark} 

\date{Received  ... / Accepted ...}

\abstract{
It has been suggested that the deficit in the number of spots on the surface of the Sun between 1790 and 1830, known as the Dalton minimum, contained an extra cycle that was not identified in the original sunspot record by Wolf. Though this cycle would be shorter and weaker than the average solar cycle, it would shift the magnetic parity of the solar magnetic field of the earlier cycles. This extra cycle is sometimes referred to as the 'lost solar cycle' or 'cycle 4b'. Here we reanalyse $^{10}$Be measurements with annual resolution from the NGRIP ice core in Greenland in order to investigate if the hypothesis regarding a lost sunspot cycle is supported by these measurements. Specifically, we make use of the fact that the Galactic cosmic rays, responsible for forming $^{10}$Be in the Earth's atmosphere, are affected differently by the open solar magnetic field during even and odd solar cycles. This fact enables us to evaluate if the numbering of cycles earlier than cycle 5 is correct. For the evaluation, we use Bayesian analysis, which reveals that the lost sunspot cycle hypothesis is likely to be correct. We also discuss if this cycle 4b is a real cycle, or a phase catastrophe, and what implications this has for our understanding of stellar activity cycles in general.
} 

\maketitle 

\section{Introduction}
It was first suggested by \citet{Loomis1870} based on auroral observations that a sunspot cycle was lost in the 1790s in the original sunspot record by Wolf -- meaning that cycle 4, which normally would have lasted from 1784 to 1799, contained two cycles: cycle 4a lasting from 1784 to 1793 and cycle 4b lasting from 1794 to 1799. This suggestion has since caused some scientific debate. Part of the reason for this is that the parity of the solar cycles does have consequences for our understanding of the solar dynamo, as the 11-year solar cycle covers only half of the 22-year magnetic polarity Hall cycle. Therefore, using various parameters, it is possible to distinguish between even and odd solar cycles, and an extra cycle would thus change the parity of cycles prior to cycle 4b.

Loomis' idea from 1870 was taken up again by \citet{2001A&A...370L..31U} who used the group sunspot number (GSN) calculated by \citet{1998SoPh..181..491H} to show that not only did a cycle 4b scenario agree better with the GSN, it also made it possible to extend the Gnevyshev-Ohl (GO) rule \citep{GO1948} to cycles before the Dalton minimum. The GO rule describes how even cycles typically have smaller amplitudes than the following odd cycles. A plausible explanation for the GO rule is that the Sun's magnetic field consists of both a dynamo and a fossil field. During odd cycles, the Sun's poloidal magnetic field is aligned with the fossil field whereas it is misaligned during even cycles \citep[see i.e.][]{2010LRSP....7....3C}. The 2001 Usoskin et al. study was followed up by \citet{2002GeoRL..29.2183U} who demonstrated how $^{10}$Be measurements from Dye 3, South Greenland, with annual resolution did not exclude the cycle 4b scenario and how the Waldmeier relation \citep[see i.e.][]{2010LRSP....7....3C} was also improved by including a lost cycle. These results were, however, called into question by \citet{2002A&A...396..235K} who reexamined both the GSN, $^{10}$Be measurements, $^{14}$C measurements from tree rings and auroral records without finding any statistical evidence for a lost cycle. \citet{2002A&A...396..235K} did not call into question the improved extension of the GO rule, but argued that this only suggested a phase shift and not a lost cycle. A reply to the criticism by \citet{2002A&A...396..235K} was provided by \citet{2003A&A...403..743U}, who argued that the negative results by \citet{2002A&A...396..235K} were mainly caused by the fact that they did not properly account for the uncertainties of the GSNs and used arithmetic averages rather than weighted averages. Finally, \citet{2009ApJ...700L.154U} claim to have resolved the mystery using recovered solar drawings from the Dalton minimum by Staudacher and Hamilton \citep{2008SoPh..247..399A, 2009AN....330..311A, 2009SoPh..255..143A}. These drawings were used to reconstruct the solar butterfly diagram during the Dalton minimum which shows the presence of high-latitude sunspots around 1793 suggesting that a new cycle started around this year.

The Dalton minimum marks a period characterised by very few or no sunspots, but in general this is not important for the lost cycle hypothesis. What is important for the lost cycle hypothesis is the GO rule, the Waldmeier relation, the butterfly diagram and the parity of the solar magnetic field. The subject of this study is the latter.
\begin{figure*}
\includegraphics[width=\textwidth]{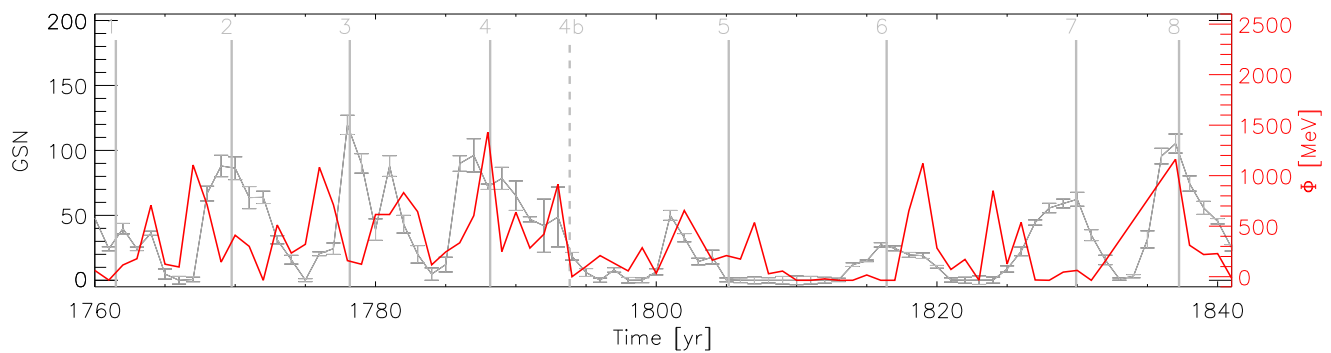}
\caption{Comparison between the observed GSN (black line and left axis) and the solar modulation potential $\Phi$ reconstructed from $^{10}$Be measurements from the NGRIP ice core on Greenland (red line and right axis). The solid grey lines indicate the midpoints of the canonical solar cycles, whereas the dashed line indicates the midpoint of cycle 4b.}
\end{figure*}
In this paper, we reanalyse $^{10}$Be measurements from the NGRIP ice core from North Greenland \citep{2009GeoRL..3611801B} in order to reevaluate the lost cycle hypothesis. The analysis  takes advantage of the so-called hysteresis effect, which has been reported by various authors \citep{1984Ap&SS.106...61M, 1995AdSpR..16..245M, 2006JApA...27..455G, 2007SoPh..245..369M, 2014a}. The hysteresis effect describes how Galactic cosmic rays (GCR) are modulated differently by the solar open magnetic field during even and odd solar cycles due to the polarity of the field. During the declining phase of even cycles and the onset of odd cycles, when the polarity is positive, the GCRs will experience an inward drift over the heliographic poles and an outward drift at equator, while the opposite will be the case when the polarity is negative. The outward drift will reduce the rate at which the GCR moves inwards, whereas an inwards drift will increase it. All things being equal, this should not make a difference in the solar modulation of GCRs, but Ulysses observations have suggested that the drifts at equator are likely to be more significant than the drifts at the poles. \citep[see i.e.][and reference herein for a full description of this effect.]{2000GeoRL..27.2453V}. This means that the flux of GCR particles at the Earth's orbit will recover faster from the strong open solar magnetic field associated with cycle maxima after even cycles compared to odd cycles. For odd cycles, this can be observed as a delay of up to two years between changes in the open solar magnetic fields and the flux of GCRs at the Earth's orbit (the hysteresis effect), while the relation is linear for the even cycles -- i.e. no delays \citep{2014a}.
 
\citet{2014b} used the hysteresis effect to make an improved reconstruction of the solar modulation strength based on $^{10}$Be measurements from NGRIP. In this study, a parametric form of the ellipse equation was used to model the hysteresis effect and the results were then used in the physical model of \citet{2002JGRA..107.1374U, 2004A&A...413..745U} to evaluate the long-term trend (cycle-to-cycle) of the GSN. Here, we present a method to model the hysteresis effect using differential equations that does not cause problems when linking even and odd cycles [encountered by \citet{2014a,2014b}] and we adjust the physical model by \citet{2002JGRA..107.1374U, 2004A&A...413..745U} to make it work on a sub-cycle time scale. This makes it possible for us to compare two reconstructions of the GSN around the Dalton minimum -- one with the lost cycle and one without. One of the advantages of our evaluation of the $^{10}$Be measurements compared to the evaluations by \citet{2002GeoRL..29.2183U} and \citet{2002A&A...396..235K} is that we do not only rely on the years 1785--1805 to evaluate the hypothesis, because it is not just cycle 4 that is affected by the lost cycle in our analysis, but, as in the evaluation of the GO rule, it is all cycles around the Dalton minimum that change parity by adding an extra cycle.  

\section{Observations}
Two different sets of observations are used in this analysis - the GSN record and the solar modulation potential of GCRs calculated based on $^{10}$Be concentrations in the NGRIP ice core.

\subsection{The group sunspot number record}
The main reason for using the GSN and not the Wolf sunspot number (WSN) is that the GSN is more reliable and homogeneous than the WSN before 1849 \citep{2013LRSP...10....1U}. There are a number of reasons for this, mainly that sunspot groups are more easily identified than individual spots. Also, the GSN relies on all observers, whereas the WSN only relies on one primary observer. Finally, the method we use for calculating reliable uncertainties of the sunspot numbers is based on individual daily sunspot values, which are available for the GSN record, but not for the WSN record. The GSN record was compiled by \citet{1998SoPh..181..491H} based on the WSN and 65,941 additional observations from 117 observers active before 1874.

As the $^{10}$Be measurements only have annual resolution, we also only need the GSN calculated with annual resolution. The most simple way to calculate the annual value of the GSN and the uncertainty of this value would be to just calculate the arithmetic mean and the uncertainty as the uncertainty of the mean. Unfortunately, as showed by \citet{2003SoPh..218..295U} this is not the correct way to calculate the annual value of the GSN and its uncertainty. The main reason is that not all daily or even weekly or monthly values have the same quality and in order to take this into account, the annual values should be calculated, as weighted mean values, where the weights are calculated based on the quality of the monthly values (which again are calculated based on the quality of the weekly and daily values). Apart from this, we are left with another problem for our analysis that is related to the fact that the Bayesian analysis relies heavily on the uncertainties and especially on the uncertainties of GSNs in years with GSNs very close to zero. In a year where no sunspots are observed at all, then the annual value of the GSN will be zero, but the uncertainty of the annual value will in principle (if it is calculated as the uncertainty of the mean) also be zero -- independent of how many observations the annual value is based on. We therefore apply the recipe by \citet{2003SoPh..218..295U}, described below, for calculating the annual GSN and its uncertainty. 

As we are working with a period characterised by low sunspot numbers we assume that the monthly GSN can be calculated as the mean of the daily values. The yearly GSN is then calculated as a weighted average of the monthly GSN:
\begin{equation}
{\rm GSN}_{y}=\frac{1}{w}\sum_{m=1}^{12}{w_{m}{\rm GSN}_m},
\end{equation}
where the statistical weights are calculated as $w_m=1/\sigma_m^2$ and $w=\sum w_m$. The uncertainty of the annual $\sigma$ can now be calculated as:

\begin{equation}
\sigma = \left\{
\begin{array}{lll}
( \sigma_{actual}+\sigma_{expected} )/2 & {\rm if} & \sigma_{actual} < \sigma_{expected},\\
\sigma_{actual} & {\rm if} & \sigma_{actual} > \sigma_{expected},  
\end{array}
\right.
\end{equation}

where the expected mean uncertainty $\sigma_{expected}$ is calculated as:
\begin{equation}
\sigma_{expected}=1/\sqrt{w}
\end{equation}
and the actual mean uncertainty as:
\begin{equation}
\sigma_{actual}= \sqrt{\frac{1}{(n-1)w}\sum{w_m\left({\rm GSN}_{m}-{\rm GSN}_{y}\right)^2}}.
\end{equation}

For the months where GSN$_m$ equals zero, we have the problem that $w_m \rightarrow \infty$. As shown in \citet{2003SoPh..218..295U}, this can be solved analysing the relation between $\sigma_m$ and the number of spotless days within a given month. This relation is nearly exponential and predicts that $\sigma_m=0.51$ for 30 spotless days within a month. We thus, as in \citet{2003SoPh..218..295U}, set all values of $\sigma_m \rightarrow 0$ to $\sigma_m = 0.51$.

The calculated GSNs and the associated uncertainties are plotted in Figs.~1~\&~2.

\begin{figure*}
\includegraphics[width=\textwidth]{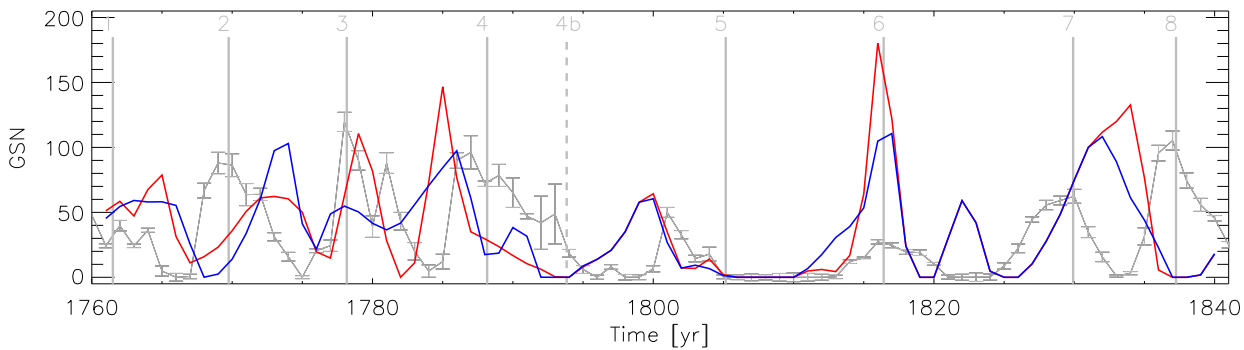}
\caption{Comparison between the observed GSN (black line), the GSN calculated based on $^{10}$Be measurements and assuming a cycle 4b (blue line) and the GSN calculated based on $^{10}$Be measurements and not assuming a cycle 4b (red line). The calculation including cycle 4b is supported by the Bayesian analysis. Again, the solid grey lines indicate the midpoints of the canonical solar cycles, whereas the dashed line indicates the midpoint of cycle 4b.}
\end{figure*}
\subsection{Modulation potential}
To our knowledge, $^{10}$Be measurements with annual resolution in the period 1750--1950 that we analyse here, are only available from the NGRIP and the Dye-3 ice cores \citep{2009GeoRL..3611801B, 1990Natur.347..164B, 1998SoPh..181..237B}. Here, the NGRIP cores are clearly superior to the Dye-3 ice cores, mainly because the NGRIP cores, due to their location on North Greenland, are less affected by climatic effects, but also, secondly, as there are indications of dating uncertainties in the older parts of the Dye-3 cores, where Dye-3 dating was established by a different method than in the newer parts \citep{2009GeoRL..3611801B}. We therefore only analyse the measurements from the NGRIP ice cores.

The solar modulation potential (or strength) $\Phi$ is a quantity that measures the solar modulation of GCR particles in the heliosphere and it can be calculated from the $^{10}$Be flux based on the recipe in \citet{2009GeoRL..3616701K}, which combines theoretical production functions with changes in the Earth's magnetic field (though these changes are negligible, but we still use the recipe for the non-linear conversion from $^{10}$Be flux to modulation potential). Note that as our $^{10}$Be record has annual and not 5-year resolution, and as we do not apply a 61-point binomial filter to the record we do have a few years with such a high $^{10}$Be flux that it corresponds to slightly negative modulation potentials (i.e. 1772, 1810, 1811, 1813, 1814, 1816, 1817, 1823, 1828 and 1831). This can be caused by uncertainties in either the theoretical production function or the $^{10}$Be measurements, or it could suggest that the flux of GCR to the heliosphere is not constant over the timescale of interest. The theoretical production functions calculated by \citet{1999JGR...10412099M} \citep[and used by][]{2009GeoRL..3616701K} using numerical models are based on comparison between sunspot numbers, neutron monitor data and cosmogenic isotopes. As sunspot numbers represent a threshold phenomenon, the zeropoints between these relationships are associated with some uncertainty and we suspect that this is the cause behind the slightly negative modulation potentials. There are two possible solutions to this, either shifting the zeropoints or truncating the negative values to zero. Based on the treshold nature of the sunspots, we use the last solution in this analysis. We have tested that the conclusion concerning the lost cycle hypotesis is insensitive to the choice of method.

We do not apply the linear interpolation to the $^{10}$Be fluxes used by \citet{2014b} for correcting the data set. The linear interpolation was applied by \citet{2014b} in order to ensure continuity of the observations, which was needed for the analysis of the hysteresis effect, not due to physical arguments. As our analysis does not need the observations to show continuity, we do not interpolate the observations. The calculated modulation potential is plotted in Fig.~1.

We emphasise that the selection criteria used to choose observations for the analysis are based on which measurements are the most reliable. The analysis could also have been performed using e.g. the WSN or the $^{10}$Be measurements from the Dye-3 ice core, but as we argue above these observations are less reliable than those used in this study, so this would, in turn, have influence the reliability of the conclusions.

\section{Analysis}
For the analysis, we need a model that can be used to calculate the GSN from the modulation potential. The model we use is based on the {\it physical model} by \citet{2004A&A...413..745U}, but we include an extra step to account for the hysteresis effect. This means that our model cannot be completely physical in the sense that we need some empirical relation in order to account for the hysteresis effect. The reason is that we, currently, do not exactly know what causes the hysteresis effect \citep[see e.g. discussion in][]{2014a}, which makes it difficult to model it. On the other hand, it is really clear from observations that the effect is real and significant \citep{2014a}, and it is straight forward to describe the observed effect mathematically -- with some constants that will then have to be constrained from observations empirically. We therefore construct a model, as described in details below, with four free parameters. These parameters are then estimated for the time period from 1850 to 1950, which defines the training period. Using these parameters, we then evaluate the lost cycle hypothesis using observations from 1750 to 1850.

\subsection{The model}
The hysteresis effect is included in the model by exchanging the step in the approach by  \citet{2004A&A...413..745U} where the open magnetic flux $F_0$ is calculated:

\begin{equation}
F_0=0.023 \cdot \bar{\Phi}^{0.9},
\end{equation}

with:

\begin{equation}
F_0=g\int_0^{\tau_h}\Phi dt,
\end{equation}

where $\Phi$ is the modulation potential and $g$ is a free parameter we call the gain. The hysteresis lag $\tau_h$ has to depend on whether we are considering an even or an odd cycle. We did test different functional forms for $\tau_h$ -- i.e. a simple step function, a triangular function or a sine function etc. and found that the best result was obtained with a smoothed step function. This means that: 

\begin{equation}
\tau_t = \left\{
\begin{array}{rl}
\tau_0 & {\rm ~if~cycle~number = odd,}\\
\tau_1 & {\rm ~if~cycle~number = even,}\\
\end{array} \right.
\end{equation} 
where $\tau_0$ and $\tau_1$ are free parameters that are estimated from the observations spanning the training period 1850 to 1950. The smoothing is done by calculating a weighted mean as:

\begin{equation}
\tau_h=\frac{\sum_{i}\tau_tw(i,h)}{\sum_{i}w(i,h)},
\end{equation}
where the weights are defined as:
\begin{equation}
w(i,h)=e^{\frac{-2\left(t_i-t_h\right)^2}{s^2}}
\end{equation}
and $s$ is the smoothing width, which is estimated along with $g$, $\tau_0$ and $\tau_1$ using the observations from the time period 1850 to 1950. 

After the $^{10}$Be has been produced in the Earth's atmosphere, the atoms become adsorbed by aerosols and stay for 1 to 2 years in the lower stratosphere \citep{1981Natur.292..825R}. This delay is not modelled in the recipe by \citet{2009GeoRL..3616701K} and has so far not been taken into account in the physical model by \citet{2004A&A...413..745U}. Here, we do not take the delay explicitly into account, but it is accounted for by $\tau_0$ and $\tau_1$. If there is no delay in the transport of $^{10}$Be in the atmosphere, $\tau_0$ should be close to zero (or one as we work with annual resolution), but because there is a delay, $\tau_0$ is larger than two. This also explains why \citet{2014b} found an ellipse relationship between $^{10}$Be measurements and GSN for both even and odd cycles, instead of an ellipse relationship for odd cycles and a linear relationship for even cycles as was found between the neutron counting rates and the GSN by \citet{2014a}. We believe reason for this is the delay of the $^{10}$Be in the atmosphere, which apply only to the $^{10}$Be measurements, not to the neutron counting rates.

Note that $F_0$ in the formulation by \citet{2004A&A...413..745U} is given in $10^{14}$ Wb. In our formulation, this factor and the 0.023 constant is included in the gain factor $g$.

In the next step, we follow \citet{2004A&A...413..745U} and use the relation from \citet{2000Natur.408..445S} to calculate the source function $S$ from the open magnetic flux:

\begin{equation}
\frac{dF_0}{dt}=S-\frac{F_0}{\tau_s},
\end{equation}
where $\tau_s = 4$ years represents the characteristic decay time of the open network flux. As noted by \citet{2004A&A...413..745U}, taking the time derivative of relatively noisy observations as the $^{10}$Be measurements causes fluctuations from one point to the next. This is of course a problem if the aim is to visually match solar cycles, but this is not the aim of this analysis. Here, the aim is to do a statical test of the lost cycle hypothesis and the statical test will account for such fluctuations. We do therefore not apply the 11-year smoothing for calculating the source functions as done by \citet{2004A&A...413..745U}

The last step is to calculate the GSN from the source term. This is done as in \citet{2004A&A...413..745U} by solving the following equation:

\begin{equation}
S({\rm GSN})=\alpha \cdot \left(24.35+22 \cdot {\rm GSN}-0.061 \cdot {\rm GSN}^2\right),
\end{equation}
where $\alpha = 1.95 \cdot 10^{11}$ Wb/yr.

\begin{figure*}
\includegraphics[width=\textwidth]{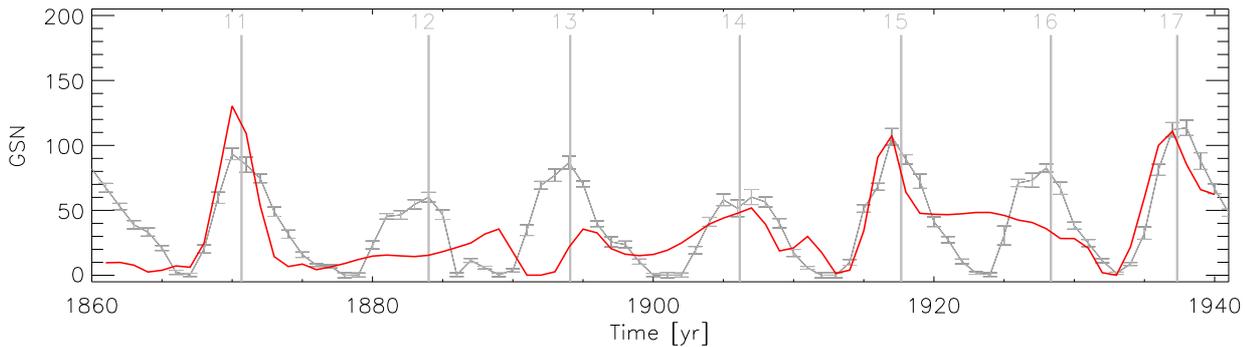}
\caption{Comparison between the observed GSN (black line) and the GSN calculated based on 10Be measurements (red line). This comparison is used to constrain the four free parameters. The solid grey lines indicate the midpoints of the canonical solar cycles.}
\end{figure*}
\subsection{The Bayesian analysis}
In order to evaluate the model, we use a Bayesian inference tool \citep[{\sc MultiNest}, ][]{2008MNRAS.384..449F, 2009MNRAS.398.1601F}. {\sc MultiNest} uses Bayes' theorem that states:

\begin{equation}
{\rm Pr}\left(\Theta \mid {\rm D},H\right) = \frac{{\rm Pr} \left( {\rm D,~}\Theta \mid H\right) {\rm Pr}\left( \Theta \mid H \right)}{{\rm Pr}\left({\rm D} \mid H\right)},
\end{equation}
where ${\rm Pr}\left(\Theta \mid {\rm D},H\right)$ is the posterior probability distribution of the parameters $\Theta$ given the observations $D$ and a model $H$, ${\rm Pr} \left( {\rm D,~}\Theta \mid H\right)$ is the likelihood, ${\rm Pr}\left( \Theta \mid H \right)$ is the prior (assumed in all cases to be uniform) and ${\rm Pr}\left({\rm D} \mid H\right) \equiv {\cal Z}$ is the Bayesian evidence. Here, the advantage compared to conventional statistical tools is that the calculated Bayesian evidence is properly normalised to the number of free parameters in the model. This is also the idea known as Occam's razor: {\it a simpler theory with compact parameter space will have a larger evidence than a more complicated one, unless the latter is significantly better at explaining the data} \citep{2008MNRAS.384..449F}. This is exactly what we want to investigate here: Does an extra sunspot cycle lead to a significantly better explanation of the relation between the $^{10}$Be measurements and the GSN?

{\sc MultiNest} is used for this problem as it is a well tested tool, which has been used for a variety of different problems in astrophysics, mainly due to the robustness and economy of this tool. We also refer to the tests we did of {\sc MultiNest} in \citet{2013ApJ...767...34K} and \citet{2014A&A...570A..41K}, where we compared {\sc MultiNest} with other maximum likelihood estimaters, these test all showed agreement between the estimated parameters and model evidences within the uncertainties. This means that the results in this study would have been the same if we had used another tool than {\sc MultiNest}.

Having calculated the Baysian evidence for a model including cycle 4b, lets call this ${\cal Z}_1$, and a model not including cycle 4b, lets call this ${\cal Z}_2$, we can calculate the {\it Bayes' factor} ${\cal K}$:

\begin{equation}
{\cal K}=\frac{{\cal Z}_1}{{\cal Z}_2}.
\end{equation}

Here, the lost cycle hypothesis is supported if ln${\cal K}$ is larger than zero and, according to \citet{Jeffreys1961}, the evidence for the hypothesis is decisive if ln${\cal K}$ is larger than 5.

In order to calculate the Baysian evidence we need to define a function for the likelihood ${\rm Pr} \left( {\rm D,~}\Theta \mid H\right)$. If we assume that the errors between the model and the GSN are given by a normal distribution \citep[see e.g.][]{2013MNRAS.430.2313C}:

\begin{equation}
f = \prod_{i=1}^{n} \frac{1}{\sqrt{2\pi}\sigma_i}e^{-\frac{\left(D_i-H_i \right)^2}{2\sigma_i^2}},
\end{equation}
where $D$ represents the observations (the GSN), $H$ is the model and $\sigma$ is the uncertainty of $D$, then we obtain the following logarithmic likelihood function:

\begin{equation}
\ell =  {\rm ln} f = -\sum_{i=1}^{N}{\rm ln}\sqrt{2\pi}\sigma_i-\frac{1}{2}\sum_{i=1}^{N}\left( \frac{{\rm ln}~D_i-{\rm ln}~H_i}{\sigma_i} \right)^2.
\end{equation}
The assumption that the errors between the model and the GSN are given by a normal distribution is based on the central limit theorem, but we stress that this is an assumption. We did test other  likelihood functions -- i.e. functions assuming $\chi^2$ distribution with two or more degrees of freedom \citep{2003A&A...412..903A}. These tests all supported the conclusions of this paper, independent of the choice of likelihood function.

In this study we use eq. 12 for two different purposes. First we use it for parameter optimisation (in section 3.3) in order to calibrate the model described in eqs. 5--11. This is done using a {\it control period} (1850--1950). Then we use it for model selection (in section 3.4) in order to evaluate the lost cycle hypothesis. This is done using observations from the {\it evaluation period} (1750-1850). The use of a control period that is different from the evaluation period for the calibration of the model ensures that free parameters are not chosen in a way that can influence the model evaluation.

\subsection{Calibration of the Model}
In order to use the model to evaluate the lost cycle hypothesis we need to calibrate the model. This means that we need to estimate the four free parameters $\tau_0$, $\tau_1$, $s$ and $g$ in the model (see eq. 2--4). This is done by using the modulation potential (calculated from the $^{10}$Be measurements) as input to the model in order to reproduce the GSN for the period 1850--1950. The other free parameters in the model, apart from $\tau_0$, $\tau_1$, $s$ and $g$, are the midpoints of the sunspot cycles during the period. Here, the first guess was taken as the maxima of a smoothed version of the GSN and the midpoints could then vary by $\pm$2 years -- the prior of the midpoint was uniform around the midpoint $\pm$2 years. Both the estimation of the free parameters and the evaluation of the lost cycle hypothesis were very robust against how the priors on the midpoints were chosen -- i.e. neither the value of the free parameters nor the likelihood of the lost cycle hypothesis were  changed beyond the uncertainties. 

As the model is cumulative to the extent that the modelling of one cycle depends on the previous (and in fact also on the successor)  -- the model has memory, we only calculate the likelihood function for the years 1860--1940, though we do reconstruct the GSN for the whole period from 1850--1950. This, of course means that the midpoints of the first and last cycle are very poorly determined, but these are not the aim of this study anyway.

{\sc MultiNest} (or the {\sc Python} implementation of it) contains a number of optimisation parameters that we in general do not use, one exception being the parameter called {\sc sampling\_efficiency}. This parameter determines, as the name indicates, how effective the sampling should be -- i.e. how likely is it that a given parameter set, with a given likelihood, will be considered for the posterior probability distribution. This parameter is equivalent to the cooling term in simulated annealing or the acceptance rate in Markov chain Monte Carlo methods. For the calibration of the model -- where the aim is to constrain the free parameters, a sampling efficiency of 0.3 is chosen, whereas we choose a sampling efficiency of 0.9 for the evaluation of the lost cycle hypothesis where the aim is evidence evaluation \citep[see discussion in][on how to choose the sampling efficiency]{2009MNRAS.398.1601F}. 

Fig.~3 shows the modelled GSN based on the NGRIP $^{10}$Be measurements for the period 1850--1950. For the four free parameters we obtained the following results: $\tau_0=2.54_{-0.11}^{+0.18}$ years$, \tau_1=4.58_{-0.46}^{+0.39}$ years, $s= 1.8_{-0.5}^{+0.7} $ days and $g=0.062_{+0.003}^{-0.003}\cdot10^{14}$ Wb/MeV$^2$.

\begin{figure*}
\includegraphics[width=\textwidth]{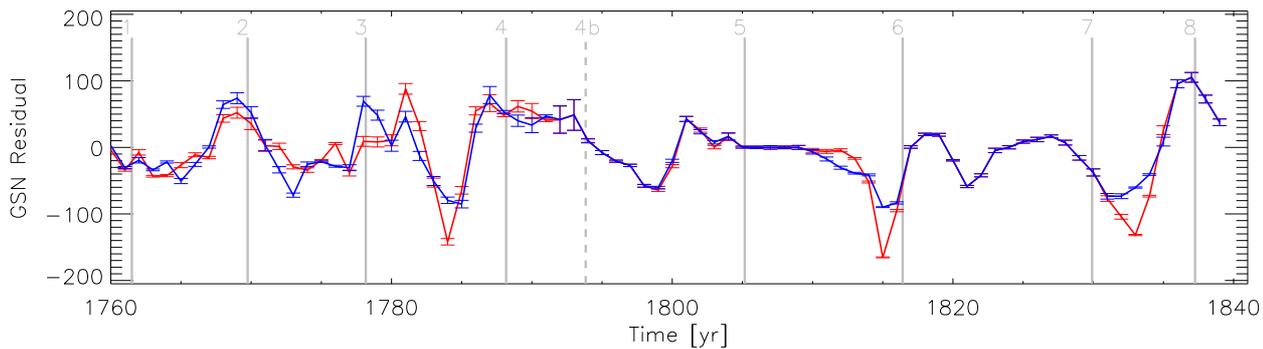}
\caption{Residuals between the observed GSN and the GSN calculated based on $^{10}$Be measurements and assuming a cycle 4b (blue line) and the GSN calculated based on $^{10}$Be measurements and not assuming a cycle 4b (red line) as in Fig. 2. Again, the solid grey lines indicate the midpoints of the canonical solar cycles, whereas the dashed line indicates the midpoint of cycle 4b.}
\end{figure*}

\subsection{Evaluation of the Lost Cycle Hypothesis}
For the evaluation of the lost cycle hypothesis, we run two models that are identical except for one point: one of them (model 1) includes an extra cycle, 4b. We use the time period 1750--1850 for the evaluation, which includes cycles 0--9. The uniform priors are given in Table~1 and a sampling efficiency of 0.9 was used. The GSN with the models overlaid is shown in Fig.~2 and scatterplots are shown in Fig.~4.

\begin{table}
\caption{Midpoints of cycles}             
\label{table:1}      
\centering                                    
\begin{tabular}{c c c c}         
\hline\hline                       
Cycle nr. & Prior [yr] & model 1 [yr] & model 2 [yr]\\    
\hline                                  
   0	& 1750.0$\pm$2.0	&	1750.0	$\pm$1.1		&	1750.0$\pm$1.1\\
   1	& 1761.5$\pm$2.0	&	1761.4	$\pm$1.1		&	1761.5$\pm$1.1\\
   2	& 1769.8$\pm$2.0	&	1768.7	$\pm$0.6		&	1770.8$\pm$0.7\\
   3	& 1778.4$\pm$2.0	&	1778.8	$\pm$1.0		&	1777.4$\pm$0.8\\
   4	& 1786.2$\pm$2.0	&	1784.9	$\pm$0.6		&	1786.4$\pm$1.1\\
   4b	& 1795.0$\pm$2.0	&	1796.2	$\pm$0.5		&	--\\
   5	&1805.2$\pm$2.0	&	1805.0	$\pm$1.1		&	1805.0$\pm$1.1\\
   6	& 1816.4$\pm$2.0	&	1814.8	$\pm$0.3		&	1814.9$\pm$0.3\\
   7	& 1829.9$\pm$2.0	&	1828.6	$\pm$0.5		&	1828.6$\pm$0.5\\
   8	& 1837.3$\pm$2.0	&	1837.8	$\pm$0.7		&	1837.9$\pm$0.7\\
   9	& 1848.2$\pm$2.0	&	1848.2	$\pm$1.1		&	1847.7$\pm$1.1\\
\hline                                           
\end{tabular}
\end{table}

\section{Results}
The results of the modelling are shown in Fig.~2, where the model including cycle 4b is shown in blue and the model without is shown in red. 

We obtain the following evidence and Bayes' factor:

\begin{equation}
{\rm ln} {\cal K} = {\rm ln}{\cal Z}_1-{\rm ln}{\cal Z}_2 = 620.3\pm0.2-614.0\pm0.2=6.3\pm0.3
\end{equation}
corresponding to a $p$-value of $99.82\pm0.06$\%, which implies that the lost cycle hypothesis is decisively supported by the $^{10}$Be measurements from the NGRIP ice core.

\section{Discussion}
Though we have shown that the lost cycle hypothesis is decisively supported by the $^{10}$ Be measurements, there are still a number of explanations that cannot be ruled out. What we can conclude, assuming that the hysteresis effect was working during the Dalton minimum as it has been working for the last 60 years, where we have continuous monitoring of GCRs with neutron counting monitors, is that it is decisively more likely that the Sun possessed and extra cycle 4b than it did not.

Another possible explanation for the Dalton minimum is the phase catastrophe scenario \citep{1994SoPh..151..351K}. Here, the idea is that the sunspot numbers can be explained as a low-dimensional chaotic system with a periodical 11-year component. This system is unstable and can, under the right conditions, drop into a laminar low-activity stage. Grand minima represent the lowest of these stages. The change from a chaotic to a laminar low system will happen through a so-called phase catastrophe, where the falling branch of a cycle will be extended thereby breaking the phase preservation. The term {\it phase} is here defined as the fraction of a cycle that has passed, and not the starting point of the cycle. \citet{2013LRSP...10....1U} provides a nice review of the low-dimensional chaotic system idea and the criticism of it. In particular Usoskin notes that different models of the sunspot numbers based on low-dimensional chaotic systems provide rather different results and that the general problem is that the sunspot record is too short to allow a proper determination of the parameters in the low-dimensional chaotic system. To this it could be added that the concept of low-dimensional chaotic systems is a purely mathematical concept -- it provides no direct explanation of the physical processes leading to the variability in the sunspot record -- except that the processes must be chaotic. Nevertheless using the phase catastrophe scenario to explain the onset of the Dalton minimum is tempting as it is capable of reproducing a number of the observed features. Along these lines, it is not clear if the parity of the cycles would be preserved during a phase catastrophe. If they are not preserved, it would explain both our results and the results related to the GO rule by \citet{2001A&A...370L..31U} without the need to include an extra cycle.

The phase catastrophe scenario also becomes particularly interesting in the light of the recent extended minimum between solar cycle 23 and 24, which has some characteristics similar to the beginning of the Dalton minimum. On the other hand, it is clear that the parity of the global solar magnetic field has changed as normal between cycle 23 and 24. Whether or not cycle 24 will be followed by 2--3 low-amplitude cycles, as cycle 4 was, is still to be revealed.

Another possible explanation for the Dalton minimum is that the configuration of the dynamo changed from a dipolar to a quadrupolar configuration during this period \citep{2013ASPC..478..167S}. Dynamo theory does suggest this as a valid explanation for grand minima \citep{2008MNRAS.388..416M} and recent analysis of helioseismic observations suggest that it is likely that the Sun contains both a dipolar and a quadrupolar dynamo mode, where the dipolar mode is responsible for the 11-year cycle and the quadrupolar mode is responsible for the biannual variability \citep{2012A&A...539A.135S, 2013ApJ...765..100S}. In this scenario, the hysteresis effect as we know it would no longer work during the Dalton minimum as the hysteresis effect relies on a dipolar configuration of the solar open magnetic field. We do not know how the modulation of GCR would be for a quadrupolar configuration of the open solar magnetic field, but for most of the possible explanations of the hysteresis effect \citep[see e.g. discussion in][]{2014a}, the modulation would be different and the hysteresis effect would therefore most likely not apply. On the other hand, if a phase catastrophe does not preserve the parity of the cycles, then the hysteresis effect would most likely not apply in neither of the scenarios.

A possible stellar connection to these questions comes from the study of the star $\epsilon$ Eridani by \citet{2013ApJ...763L..26M} where the authors identified both a 3-year and a 13-year activity cycle in a re-analysis of both archive and new Ca HK activity measurements. In the record, which extends from 1962 to 2013, they also identify a possible {\it Maunder minimum-like state for the short cycle} during the early 1990s. An interesting feature of this possible Maunder minimum-like state, is that it takes place just after a cycle with an extended falling branch -- as suggested in the phase catastrophe scenario. In other words, $\epsilon$ Eridani might be showing the same phenomenon as the Sun was showing during the Dalton minimum and it is also not clear if this is due to an extra low-amplitude cycle or to a phase catastrophe.

Progress in our understanding of both grand minima and extended cycle minima would thus be possible if we could either identify a way to measure the parity of the stellar cycles observed in other stars, mainly from the Mount Wilson \citep{1995ApJ...438..269B} and Lowell \citep{2007AJ....133..862H} observatories, or if we could identify a way to test if the phase is preserved in the Sun over grand minima like the Maunder minimum -- i.e. if all even cycles are followed by odd cycles.

\begin{acknowledgements}Funding for the Stellar Astrophysics Centre is provided by
the Danish National Research Foundation (Grant agreement no.: DNRF106). The projects has been supported by the Villum Foundation. 
\end{acknowledgements}


\begin{thebibliography}{99}
\bibitem[Van Allen(2000)]{2000GeoRL..27.2453V} Van Allen, J.~A.\ 2000, \grl, 27, 2453 
\bibitem[Appourchaux(2003)]{2003A&A...412..903A} Appourchaux, T.\ 2003, \aap, 412, 903
\bibitem[Arlt(2008)]{2008SoPh..247..399A} Arlt, R.\ 2008, \solphys, 247, 399 
\bibitem[Arlt(2009a)]{2009AN....330..311A} Arlt, R.\ 2009, Astronomische Nachrichten, 330, 311
\bibitem[Arlt(2009b)]{2009SoPh..255..143A} Arlt, R.\ 2009, \solphys, 255, 143 
\bibitem[Baliunas et al.(1995)]{1995ApJ...438..269B} Baliunas, S.~L., Donahue, R.~A., Soon, W.~H., et al.\ 1995, \apj, 438, 269 
\bibitem[Beer et al.(1990)]{1990Natur.347..164B} Beer, J., Blinov, A., Bonani, G., Hofmann, H.~J., \& Finkel, R.~C.\ 1990, \nat, 347, 164 
\bibitem[Beer et al.(1998)]{1998SoPh..181..237B} Beer, J., Tobias, S., \& Weiss, N.\ 1998, \solphys, 181, 237 
\bibitem[Berggren et al.(2009)]{2009GeoRL..3611801B} Berggren, A.-M., Beer, J., Possnert, G., et al.\ 2009, \grl, 36, 11801 
\bibitem[Charbonneau(2010)]{2010LRSP....7....3C} Charbonneau, P.\ 2010, Living Reviews in Solar Physics, 7, 3 
\bibitem[Corsaro et al.(2013)]{2013MNRAS.430.2313C} Corsaro, E., Fr{\"o}hlich, H.-E., Bonanno, A., et al.\ 2013, \mnras, 430, 2313
\bibitem[Feroz \& Hobson(2008)]{2008MNRAS.384..449F} Feroz, F., \& Hobson, M.~P.\ 2008, \mnras, 384, 449 
\bibitem[Feroz et al.(2009)]{2009MNRAS.398.1601F} Feroz, F., Hobson, M.~P., \& Bridges, M.\ 2009, \mnras, 398, 1601
\bibitem[Gnevyshev \& Ohl(1948)]{GO1948} Gnevyshev, M.~N. \& Ohl, A.~I.1948, Astron. Zh., 25(1), 18
\bibitem[Gupta et al.(2006)]{2006JApA...27..455G} Gupta, M., Mishra, V.~K., \& Mishra, A.~P.\ 2006, Journal of Astrophysics and Astronomy, 27, 455 
\bibitem[Hall et al.(2007)]{2007AJ....133..862H} Hall, J.~C., Lockwood, G.~W., \& Skiff, B.~A.\ 2007, \aj, 133, 862 
\bibitem[Hoyt \& Schatten(1998)]{1998SoPh..181..491H} Hoyt, D.~V., \& Schatten, K.~H.\ 1998, \solphys, 181, 491 
\bibitem[Inceoglu et al.(2014a)]{2014a} Inceoglu, F., Knudsen, M.~F., Karoff, C., \& Olsen, J.\ 2014a, \solphys, 289, 1387 
\bibitem[Inceoglu et al.(2014b)]{2014b} Inceoglu, F., Knudsen, M.~F., Karoff, C., \& Olsen, J.\ 2014b, \solphys, 289, 4377
\bibitem[Jeffreys(1961)]{Jeffreys1961} Jeffreys, H.\ 1961, Theory of Probability, 3rd edn. (Oxford University Press)
\bibitem[Kallinger et al.(2014)]{2014A&A...570A..41K} Kallinger, T., De Ridder, J., Hekker, S., et al.\ 2014, \aap, 570, AA41
\bibitem[Karoff et al.(2013)]{2013ApJ...767...34K} Karoff, C., Campante, T.~L., Ballot, J., et al.\ 2013, \apj, 767, 34
\bibitem[Knudsen et al.(2009)]{2009GeoRL..3616701K} Knudsen, M.~F., Riisager, P., Jacobsen, B.~H., et al.\ 2009, \grl, 36, 16701 
\bibitem[Kremliovsky(1994)]{1994SoPh..151..351K} Kremliovsky, M.~N.\ 1994, \solphys, 151, 351
\bibitem[Krivova et al.(2002)]{2002A&A...396..235K} Krivova, N.~A., Solanki, S.~K., \& Beer, J.\ 2002, \aap, 396, 235
\bibitem[Loomis(1870)]{Loomis1870}Loomis, E. 1870, Am. J. Sci., 2nd Ser, 50, 153
\bibitem[Marmatsouri et al.(1995)]{1995AdSpR..16..245M} Marmatsouri, L., Vassilaki, A., Mavromichalaki, H., \& Petropoulos, B.\ 1995, Advances in Space Research, 16, 245 
\bibitem[Masarik \& Beer(1999)]{1999JGR...10412099M} Masarik, J., \& Beer, J.\ 1999, \jgr, 104, 12099 
\bibitem[Mavromichalaki et al.(2007)]{2007SoPh..245..369M} Mavromichalaki, H., Paouris, E., \& Karalidi, T.\ 2007, \solphys, 245, 369 
\bibitem[Mavromichalaki \& Petropoulos(1984)]{1984Ap&SS.106...61M} Mavromichalaki, H., \& Petropoulos, B.\ 1984, \apss, 106, 61 
\bibitem[Metcalfe et al.(2013)]{2013ApJ...763L..26M} Metcalfe, T.~S., Buccino, A.~P., Brown, B.~P., et al.\ 2013, \apjl, 763, L26 
\bibitem[Moss et al.(2008)]{2008MNRAS.388..416M} Moss, D., Saar, S.~H., \& Sokoloff, D.\ 2008, \mnras, 388, 416 
\bibitem[Raisbeck et al.(1981)]{1981Natur.292..825R} Raisbeck, G.~M., Yiou, F., Fruneau, M., et al.\ 1981, \nat, 292, 825 
\bibitem[Simoniello et al.(2012)]{2012A&A...539A.135S} Simoniello, R., Finsterle, W., Salabert, D., et al.\ 2012, \aap, 539, A135 
\bibitem[Simoniello et al.(2013a)]{2013ApJ...765..100S} Simoniello, R., Jain, K., Tripathy, S.~C., et al.\ 2013a, \apj, 765, 100
\bibitem[Simoniello et al.(2013b)]{2013ASPC..478..167S} Simoniello, R., Jain, K., Tripathy, S.~C., et al.\ 2013b, Astronomical Society of the Pacific Conference Series, 478, 167 
\bibitem[Solanki et al.(2000)]{2000Natur.408..445S} Solanki, S.~K.,Sch{\"u}ssler, M., \& Fligge, M.\ 2000, \nat, 408, 445 
\bibitem[Usoskin et al.(2001)]{2001A&A...370L..31U} Usoskin, I.~G., Mursula, K., \& Kovaltsov, G.~A.\ 2001, \aap, 370, L31
\bibitem[Usoskin et al.(2002a)]{2002JGRA..107.1374U} Usoskin, I.~G., Mursula, K., Solanki, S.~K., Sch{\"u}ssler, M., \& Kovaltsov, G.~A.\ 2002a, Journal of Geophysical Research (Space Physics), 107, 1374
\bibitem[Usoskin et al.(2002b)]{2002GeoRL..29.2183U} Usoskin, I.~G., Mursula, K., \& Kovaltsov, G.~A.\ 2002b, \grl, 29, 2183 
\bibitem[Usoskin et al.(2003a)]{2003A&A...403..743U} Usoskin, I.~G., Mursula, K., \& Kovaltsov, G.~A.\ 2003a, \aap, 403, 743 
\bibitem[Usoskin et al.(2003b)]{2003SoPh..218..295U} Usoskin, I.~G., Mursula, K., \& Kovaltsov, G.~A.\ 2003b, \solphys, 218, 295 
\bibitem[Usoskin et al.(2004)]{2004A&A...413..745U} Usoskin, I.~G., Mursula, K., Solanki, S., Sch{\"u}ssler, M., \& Alanko, K.\ 2004, \aap, 413, 745
\bibitem[Usoskin et al.(2009)]{2009ApJ...700L.154U} Usoskin, I.~G., Mursula, K., Arlt, R., \& Kovaltsov, G.~A.\ 2009, \apjl, 700, L154 
\bibitem[Usoskin(2013)]{2013LRSP...10....1U} Usoskin, I.~G.\ 2013, Living Reviews in Solar Physics, 10, 1 
\end{thebibliography}
\end{document}